\documentclass[11pt,a4paper]{article}
\usepackage[a4paper, margin=1in]{geometry}

\usepackage{graphicx}%
\usepackage[hidelinks]{hyperref}
\usepackage{url}

\usepackage[T1]{fontenc}%
\usepackage{textcomp}%
\usepackage{newtxtext,newtxmath}%

\usepackage{multirow}%
\usepackage{natbib}%

\usepackage{caption}%
\newcommand{\figref}[1]{\figurename \ref{#1}}
\newcommand{\tabref}[1]{\tablename~\ref{#1}}

\begin{document}

\title
{Exploring the applicability of Large Language Models\\to citation context analysis}

\author{
  Kai NISHIKAWA\textsuperscript{1,2}
  \thanks{Email: \texttt{knishikawa@slis.tsukuba.ac.jp}}
  \hspace{20mm}
  Hitoshi KOSHIBA\textsuperscript{2}
}

\date{
    \small
    \textsuperscript{1} Institute of Library, Information and Media Science, University of Tsukuba\\
    \textsuperscript{2} National Institute of Science and Technology Policy (NISTEP)
}

\maketitle

\abstract{
Unlike traditional citation analysis ---which assumes that all citations in a paper are equivalent--- citation context analysis considers the contextual information of individual citations.
However, citation context analysis requires creating large amounts of data through annotation, which hinders the widespread use of this methodology.
This study explored the applicability of Large Language Models (LLMs) ---particularly ChatGPT--- to citation context analysis by comparing LLMs and human annotation results.
The results show that the LLMs annotation is as good as or better than the human annotation in terms of consistency but poor in terms of predictive performance. 
Thus, having LLMs immediately replace human annotators in citation context analysis is inappropriate. 
However, the annotation results obtained by LLMs can be used as reference information when narrowing the annotation results obtained by multiple human annotators to one, or LLMs can be used as one of the annotators when it is difficult to prepare sufficient human annotators.
This study provides basic findings important for the future development of citation context analyses.
}

\subsubsection*{keywords:}
\hspace{-3mm}
\texttt{
Scientometrics, 
Citation Context Analysis, 
Annotation,\\
Large Language Model (LLM), 
ChatGPT
}

\clearpage

\section{Introduction}
Quantitative analysis focusing on citation relationships among papers assumes that all citations are essentially and implicitly equivalent \citep[e.g.,][]{Bornmann2008, Ding2014, Lin2018}. 
In contrast, citation context analysis has been proposed to consider the contextual information of individual citations, such as the location of the citation and the semantic content of the text containing the citation. 
Although citation context analysis is expected to provide complementary findings to the traditional quantitative citation analysis, it has the drawback that the cost of creating the data necessary for the analysis is significant. 
Therefore, it is difficult to conduct studies that require a large amount of data, for example, analyzing differences in citation context trends among multiple disciplines.

In the citation context analysis, data are created by determining the contextual characteristics of each citation using the text surrounding the citation in the citing paper. 
There are two ways to create data: manual data processing, in which a human annotator manually creates data, and automatic data processing, in which data are created using machine learning and other techniques \citep{Tahamtan2019}. 
However, because the latter method often uses supervised learning---which requires training data---it is necessary to create large datasets using human annotators. 
The high cost of this annotation work is an obstacle to the development of citation context analysis.

However, with the recent development of Large Language Models (LLMs) such as GPT \citep{Brown2020}, some studies have attempted to perform general annotation tasks on behalf of human annotators \citep[e.g.,][]{Gilardi2022, He2023, Pangakis2023, Reiss2023, Rytting2023}. 
These studies clarify that LLMs sometimes outperform human annotators hired through crowdsourcing and can produce more data in a more time- and cost-efficient manner. 
However, the performance of LLMs annotation varies depending on the specific task, even for the same text classification, and it is not necessarily clear whether LLMs can immediately automate the annotation process.

To the best of our knowledge, no study has focused on whether LLMs can substitute for human annotators in scientific papers. 
Because a scientific paper is a specialized text with its own formatting and writing style and contains a large amount of specialized terminology, annotations seem different from general annotations that can be easily crowdsourced, as focused on in previous studies. 
In fact, in citation context analysis, annotations are often performed by researchers or graduate students employed as research assistants (RA), who are accustomed to reading articles and are required to be familiar with a schema and manual for annotation through a certain amount of training. 
Therefore, it is unclear whether the findings of previous studies can be applied to the citation context analysis.

This study aims to explore the applicability of LLMs to citation context analysis. 
Specifically, we will examine the following by having LLMs perform annotation tasks similar to those performed by human annotators in \citet{Nishikawa2023}, a previous study on citation context analysis:
\begin{enumerate}
    \item Can LLMs replace humans for annotations in citation context analysis?
    \item How can LLMs be effectively utilized in citation context analysis?
\end{enumerate}

The results of this study indicate that the annotation results of LLMs are comparable to or better than those of humans in terms of consistency but poor in terms of predictive performance. 
Therefore, it is not appropriate to allow current LLMs to perform annotations associated with citation context analysis on behalf of humans. 
However, the annotation results obtained by LLMs can be used as reference information when narrowing the annotation results obtained by multiple human annotators to one, or LLMs can be used as one of the annotators if securing a sufficient number of human annotators is difficult. 
This study provides basic findings important for the future development of citation context analyses.

In the following section, we provide a literature review followed by the methods and results of the experiments. 
Subsequently, based on the experimental results, we discuss whether LLMs can replace human annotators. 
Next, we examine whether LLMs can be applied to citation context analysis beyond replacing humans. 
Finally, we present our conclusions.

\section{Literature Review}

\subsection{Citation Context Analysis}
Citation context analysis is also referred to as citation content analysis. 
Although some studies distinguish between both \citep[e.g.,][]{Tahamtan2019}, we use the term citation context analysis as including citation content analysis. 
When conducting citation context analysis, the first step is to set a schema that defines the categorization of citations. 
Categories and possible values (also called classes) for each category are often set arbitrarily by researchers according to their research purposes. 
\citet{Zhang2013} divided these into syntactic and semantic categories, with the former represented by citation location and the latter by citation purpose (also called citation function or citation motivation) and citation sentiment.

Next, a dataset is created by classifying the citations to be analyzed based on the schema. 
This stage corresponds to a task known as annotation, coding, or citation classification. 
As mentioned previously, there are two dataset creation methods: human annotators (coders), machine learning, and other techniques \citep{Tahamtan2019}. 
The former is a costly method in terms of both time and money, whereas supervised learning is often used in the latter, especially for semantic categories that require human annotation \citep{Iqbal2021}. 
In addition, the distribution of classes has been reported to be highly skewed for many categories \citep[e.g.,][]{Kunnath2023-rl, Nishikawa2023}, which is another reason for the need for larger datasets for analysis.

Finally, the created dataset is analyzed for individual research purposes. 
However, because both human and machine methods require costly human annotations, studies requiring large datasets, such as comparisons of citation relationships among multiple disciplines, are not well developed. 
The few exceptions that make inter-discipline comparisons are achieved by limiting the number of categories, topics, or disciplines focused on \citep{Chang2013, Lin2018, Zhang2021, Wang2021, Nishikawa2023}. 
In other words, the cost of annotation must be reduced to allow a more flexible or large-scale study design for citation context analysis.

\subsection{Annotation by LLMs}
Annotation refers to the text classification in natural language processing tasks. 
Many previous studies have compared the results of multiple models and versions of LLMs to evaluate their performance in text classification  \citep[cf.,][]{Chang2023-jq}. 
Several studies have focused on the potential of LLMs as substitutes for annotation, comparing the annotation results obtained by multiple models and versions of LLMs \citep{He2023, Kuzman2023-qr, Pangakis2023, Reiss2023}, or more directly comparing human and LLMs annotations \citep{Gilardi2022, He2023, Rytting2023}. 
The specific tasks performed in these studies are as diverse as classifying topics for social networking posts \citep{Gilardi2022}, classifying websites as news \citep{Reiss2023}, and classifying article genres for news article headlines \citep{Rytting2023}. 
In studies comparing human and LLM annotations, several texts that are the subject of these tasks are common and do not require expertise to read. 
Therefore, crowd workers are often employed as human annotators in addition to trained annotators, such as graduate students, when making comparisons between humans and LLMs.

Although the cost of annotation is significantly lower with LLMs than with crowd workers \citep{Gilardi2022}, the question remains as to whether the quality of annotation results with LLMs is high enough to be used for analysis. 
The results of annotation tasks are often evaluated in terms of their consistency (also known as reliability) and prediction performance. 
Inter-coder agreement is often used as a consistency metric, whereas accuracy or F1 is often used as a performance metric. 
The findings of previous studies indicate that LLMs are superior or comparable to human annotators, particularly cloud workers, in terms of consistency and performance. 
Meanwhile, it has also been noted that the consistency and performance of LLMs annotation work can vary depending on the attributes of texts and categories \citep{Pangakis2023, Reiss2023}. 
Thus, no consensus has been reached on whether current LLMs, such as ChatGPT, can immediately replace human annotators.

Additionally, few studies have applied LLMs to citation classification. 
\citet{Zhang2023-ma} argued that LLMs can contribute to citation context analysis by automating citation classification; however, the applicability of LLMs classification is currently unclear and should be addressed in the future. 
\citet{Kunnath2023-rl} compared the performance of LLMs citation classification when performing parameter updating using multiple methods on the public datasets ACL-ARC \citep{Jurgens2018-nu} and ACT2 \citep{Nambanoor_Kunnath2022-tu}. 
The results show high performance when using some of the methods and that the zero-shot performance of GPT3.5 is high when targeting multiple fields (ACT2) but low when targeting a single field (ACL-ARC). 
However, \citet{Kunnath2023-rl} did not compare human annotators to LLMs, nor did it focus on the applicability of LLM-generated data for citation context analysis.

To the best of our knowledge, it is not yet clear whether LLMs can replace humans in annotating paper, a special type of text that requires expertise in reading and understanding. 
In other words, it remains to be seen whether LLMs can be used in applied citation context analysis research, in which researchers create and analyze data that categorize individual citations for their own research purposes.

\section{Methods}

\subsection{Task and Data}
Many categories are used in citation context analysis, and annotations vary widely according to the categories used. 
We let LLMs perform annotation on the two categories used in \citet{Nishikawa2023}: citation purpose and citation sentiment. 
Descriptions are presented in \tabref{tab: Citation purpose and citation sentiment}.

\begin{table}[htbp]
    \centering
    \caption{Citation purpose and citation sentiment}
    \label{tab: Citation purpose and citation sentiment}
    \begin{tabular}{rp{80mm}}
        \hline
        Categories & Description \\
        \hline
        \hline
        \textbf{Citation Purpose} &
        The type of purpose for citing a cited paper \\&
        \textbf{5 classes:}
        1. Background, 2. Comparison, 3. Criticize, 4. Evidence, 5. Use \\
        \textbf{Citation Sentiment} & 
        The mental attitude of the author of a citing paper when citing a cited paper 
        \\&
        \textbf{3 classes:} 
        1. Positive, 2. Negative, 3. Neutral \\
        \hline
    \end{tabular}
\end{table}

This study focuses on the same tasks as those in \citet{Nishikawa2023} for the following three reasons. 
First, \citet{Nishikawa2023} simply organizes categories and their classes (also called values) after reviewing previous studies on citation context analysis. 
Second, the manual used for human annotators is publicly available and can be annotated using LLMs under the same conditions as humans. 
Third, because the gold standard data used in the analysis are publicly available \citep{Nishikawa2023a}, the predictive performance of the annotation results from the LLMs can be evaluated. 
Regarding the last reason, in \citet{Nishikawa2023}, the data used in the analysis were prepared using the following procedure:

\begin{enumerate}
    \item Two annotators, a researcher and a graduate student employed as a research assistant, independently annotated all data according to the manual.
    \item After the initial annotation, the annotators explained why they had determined a value for a citation in which the results did not match\footnote{This phase is called ``discussion'' \citep{Lin2018}.}.
    \item Finally, data from the annotator with the lowest number of value modifications for each category were used in the analysis.
\end{enumerate}

While \citet{Nishikawa2023} set six categories, this study focuses only on citation purpose and sentiment for the following reasons. 
First, citation purpose and sentiment are the main categories addressed in many previous studies that have conducted citation context analyses \citep{Lyu2021}. 
In addition, automating their annotations is difficult because it is necessary to understand the semantic content of the surrounding text in which the target-cited paper is mentioned to determine the class. 
However, the classification of other categories that correspond to syntactic categories \citep{Zhang2013} can be automated relatively easily because they can be processed without understanding the meaning of the text. 
Therefore, we believe that citation purpose and sentiment would particularly benefit from automation, and we address only these two categories in this study.

In this study, we allowed the LLMs to annotate the same texts that were the subject of annotation for citation purpose and sentiments in \citet{Nishikawa2023}. 
\citet{Nishikawa2023} created 1,174 data by human annotation for citation purpose and citation sentiment, respectively. 
However, we annotated 181 of them with LLMs for which the text of the citing papers could be obtained in the Journal Article Tag Suite (JATS) -XML format. 
We have limited the annotation target to papers in JATS-XML format because they are expected to be almost free from errors and labor involved in text extraction. 
Therefore, if LLMs annotation is successful, it can be easily applied on a large scale. 
In addition, many of the papers that are the target of annotation in \citet{Nishikawa2023} are available only in PDF format. 
In such cases, it is often difficult to accurately and automatically extract text and citation information from PDFs. 
Therefore, when evaluating the annotation results, the performance of the LLMs and the extraction accuracy of the target data must be considered. 
Manual text extraction from PDFs can avoid this problem but at a higher cost. 
For these reasons, we limited the scope of this study to articles that could be collected in JATS-XML format.

\subsection{Type of LLMs}
The LLMs annotation was performed using the API provided by OpenAPI Inc. 
The LLM model used in this study was \texttt{gpt3.5-turbo-0310}. 
The \texttt{temperature}, which is the parameter for creativity or the randomness of replies, was set to $0.5$
\footnote{According to the official explanation (\url{https://platform.openai.com/docs/api-reference/chat/create}, Last access:2023/May/06), it takes a value between 0 and 2, and defaults to 1. A value of about 0.2 always returns almost the same response, while a value of about 0.8 returns a random result. 
Considering these factors, we set an intermediate value this time.}.

\subsection{Prompts}
In LLMs, any task can be executed by providing instructions (prompts) in a natural language, and the task results can differ depending on the expression of the prompt. 
For example, to have LLMs prepare a summary of a certain paper, there are multiple patterns of possible prompts, such as 
``Please summarize the paper given below,'' 
``Please give a brief summary of the paper,'' 
and ``Please summarize the paper in about 200 words,'' 
and the results may differ among them. 
In addition to these differences in expression, using certain techniques such as few-shot \citep{brown2020language} or chain-of-thought \citep{wei2023chainofthought} can change the results\footnote{Furthermore, depending on the parameters of the LLMs and the nature of the data, the results may differ even for the same prompt}.

Therefore, in this study, we set up multiple patterns of prompts based on the manual used by human annotators in \citet{Nishikawa2023} but with almost the same content. 
Because specific prompts differ depending on the citation purpose and sentiment, the following describes the prompt patterns for each. 
The prompts are included in Online Resource 1.

\subsubsection{Prompt patterns for citation purposes}
The manual for citation contains the following elements: 
1. types of possible classes, 
2. definitions for each class, 
3. procedures for annotation, 
and 4. Keywords and example sentences based on class determination \citep{Nishikawa2023}. 
Correspondingly, in addition to the basic instructions, four patterns of prompts for citation purpose were established, including the following elements:

\begin{enumerate}
    \item Types of class only (Simple)
    \item Types of class and their definition (Basic)
    \item Types of class, their definitions, and procedures for annotation (Precise)
    \item Types of class, definitions, annotation procedures, and keywords and example sentences (Full)
\end{enumerate}

The fourth pattern (Full), which includes all elements, is almost identical to the manual used in \citet{Nishikawa2023}. 
However, the original manual included instructions that did not directly affect the annotation results, such as the handling of the files to be worked on. 
This study excluded such instructions from the prompts. 
In addition, although the original manual instructs annotators to consider ``the title of the section in which the citation in question is being made,'' in this experiment, we excluded that part of the annotation to reduce the time and effort required to extract the title.

\begin{figure}
    \centering
    \includegraphics[width=120mm]{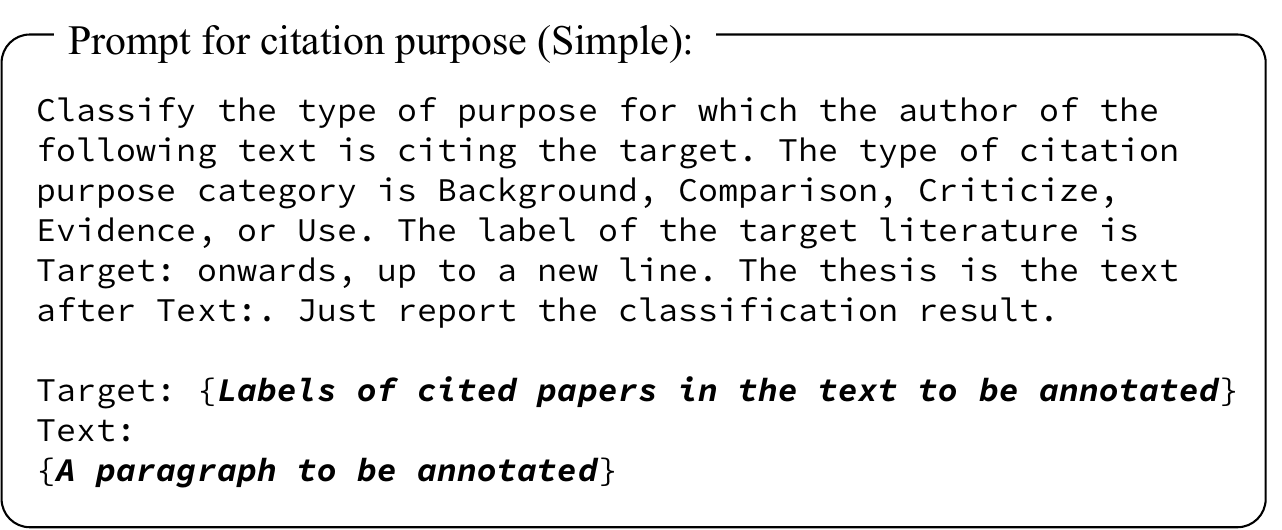}
    \caption{An example of prompt}
    \label{fig:prompt}
\end{figure}

Moreover, although the published versions of the manual and target papers were written in English, the original manual was written in Japanese. 
Therefore, eight prompt patterns were established by writing the above four patterns in Japanese and English. 
For example, the simplest pattern of prompts (Simple, EN) containing only the type of class is shown in \figref{fig:prompt}.

\subsubsection{Prompt patterns for citation sentiment}
Although the elements included in the original manual for citation sentiment are the same as those for citation purpose, there are no instructions regarding procedures for annotation. 
Therefore, three patterns of prompts are set in citation sentiment, except ``Precise,'' as follows:

\begin{enumerate}
    \item Types of class only (Simple)
    \item Types of class and their definition (Basic)
    \item Types of class, their definitions, and keywords and example sentences (Full)
\end{enumerate}

Finally, six prompt patterns were set by writing the above in Japanese and English. 
Patterns containing all the elements (Full) are generally identical to those in the original manual; however, as in the case of citation purpose, it excludes instructions that do not directly affect the annotation results contained in the original manual.

\subsection{Evaluation Metrics}
\begin{figure}
    \centering
    \includegraphics[width=120mm]{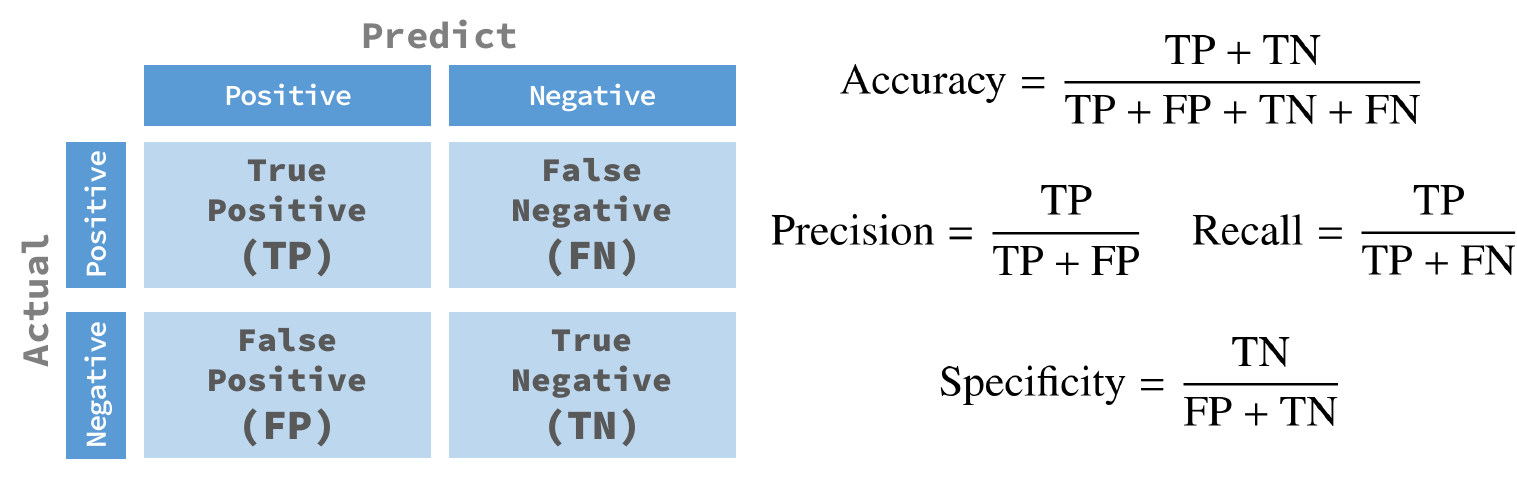}
    \caption{Definitions of indicators for performance}
    \label{fig:math}
\end{figure}

As mentioned in the Literature Review, several studies have evaluated the results of LLMs annotations in terms of consistency and predictive performance. 
These two perspectives were also used in this study, where we use \citet{Nishikawa2023a} as the gold standard. 
In addition, multiple applicable metrics exist for both perspectives. 
We used the simple agreement rate between annotators and Cohen’s kappa for consistency and the metrics shown in \figref{fig:math} for performance.

\section{Experiments}\label{sec:experience}

\subsection{Distribution of Data}
\tabref{tab:cite_pattern} and \tabref{tab:sdgs} show the distributions of the targets annotated in this study. 
In \citet{Nishikawa2023}, the analysis unit was a citation pair, which is a pair of citing papers and one of the papers cited by it, all of which were related to either renewable energy (SDG7) or climate change (SDG13). 
Both the citing and cited papers were classified as Natural Sciences (NS) or Social Sciences and Humanities (SSH), and the following four patterns of relationships between disciplines were set: NS citing NS (NS-NS), NS citing SSH (NS-SSH), SSH citing SSH (SSH-SSH), and SSH citing NS (SSH-NS). 
As mentioned in the Methods section, this study used some of these as targets for annotation. 
\tabref{tab:cite_pattern} shows the distribution of citation pairs for each of the patterns used in this study, and \tabref{tab:sdgs} shows the distribution of the citing papers by research topic, that is SDG7 or 13.

\tabref{tab:pp} and \tabref{tab:st} present the gold standards for annotation used in this study. 
In other words, they are a selection of annotation results for citation purpose and sentiments from \citet{Nishikawa2023}, and their citing papers are available in the JATS-XML format.

\begin{table}[htbp]
    \caption{Distribution of citation pairs by patterns}
    \label{tab:cite_pattern}
    \centering
    \includegraphics[width=50mm]{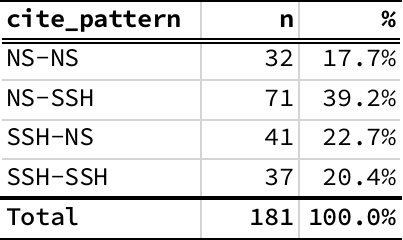}
\end{table}

\begin{table}[htbp]
    \centering
    \caption{Distribution of citing papers by topics}
    \label{tab:sdgs}
    \centering
    \includegraphics[width=42mm]{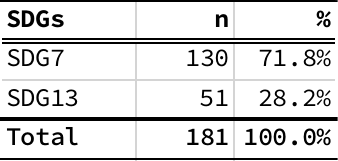}
\end{table}

\begin{table}[htbp]
    \centering
    \caption{Distribution of the gold standard on citation purpose}
    \label{tab:pp}
    \includegraphics[width=48mm]{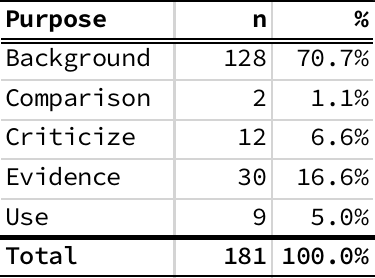}
\end{table}

\begin{table}[htbp]
    \centering
    \caption{Distribution of the gold standard on citation sentiment}
    \includegraphics[width=48mm]{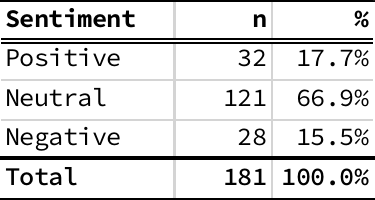}
    \label{tab:st}
\end{table}

\subsection{Consistency}
First, we compared the annotation results of the human annotators in \citet{Nishikawa2023} and ChatGPT in this study in terms of consistency. 
ChatGPT was given a prompt (Full, EN) that was nearly identical to those in the manuals used in \citet{Nishikawa2023} and was asked to annotate twice each of the citation purpose and citation sentiments for all 181 data. 
\tabref{tab:compare} shows the simple agreement rate and Cohen’s kappa for each of the annotation results by ChatGPT and the results at the time the two annotators independently annotated in \citet{Nishikawa2023}, i.e., before the ``discussion.'' 
It can be seen in \tabref{tab:compare} that ChatGPT is more consistent than humans with respect to both citation purpose and citation sentiment.

\begin{table}[htbp]
    \centering
    \caption{Comparison of consistency between humans and ChatGPT }
    \label{tab:compare}
\includegraphics[width=100mm]{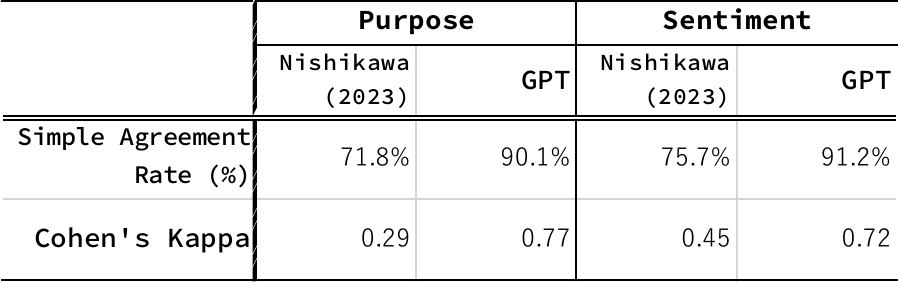}
\end{table}

Next, we compared the consistency of the annotation results using ChatGPT with all the patterns of prompts described in the Methods section. 
Using eight patterns for citation purpose and six for citation sentiment prompts, we had ChatGPT annotate all 181 data points twice each. 
\tabref{tab:pp_agree} shows the number of cases in which the results at each prompt did not agree, the simple agreement rate, and Cohen’s kappa for citation purpose.

\begin{table}[htbp]
    \centering
    \caption{Consistency by prompts for citation purpose}
    \label{tab:pp_agree}
    \includegraphics[width=80mm]{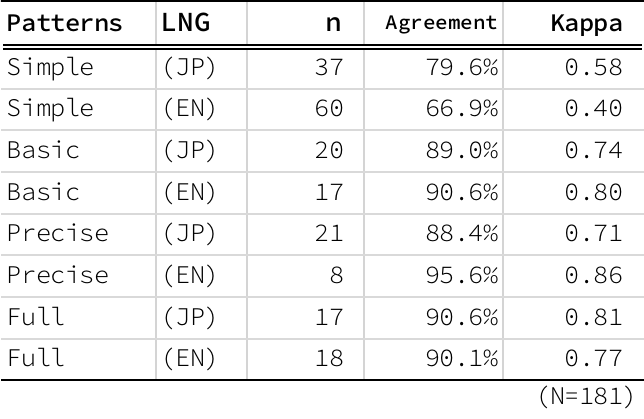}
\end{table}

As shown in \tabref{tab:pp_agree}, for citation purpose, the highest consistency was found in the prompt (Precise, EN) that provided types of classes, their definitions, and annotation procedures in English, with eight cases ($4.4\%$) differing between the first and second prompts and $95.6\%$ remaining consistent. 
The lowest consistency was for the simple prompt in English (Simple, EN), with 30 cases ($33.1\%$) differing and a simple agreement rate of $66.9\%$. 
Prompts other than this pattern exceeded the agreement rates of the human annotators. 
Interestingly, the prompt, including all elements (full), which is similar to the original manual for humans, was less consistent. 
This finding suggests that consistency does not necessarily increase with more detailed instruction.

\tabref{tab:se_agree} summarizes the consistency of the annotation results for each prompt for citation sentiment. 
The table shows that the highest consistency was for the prompt that gave only the types of classes in Japanese (Simple, JP), with one case ($0.6\%$) differing, and the lowest consistency was found in the prompt (Full, EN), with 16 cases ($8.8\%$) differing. 
However, all patterns outperform the agreement rate by human annotators, and the consistency of the prompt Simple is significantly higher than for citation purpose. 
In addition, focusing only on the prompts written in English, the more detailed and specific the instructions, the less consistent they become.

\begin{table}[htbp]
    \centering
    \caption{Consistency by prompts for citation sentiment}
    \label{tab:se_agree}
    \includegraphics[width=80mm]{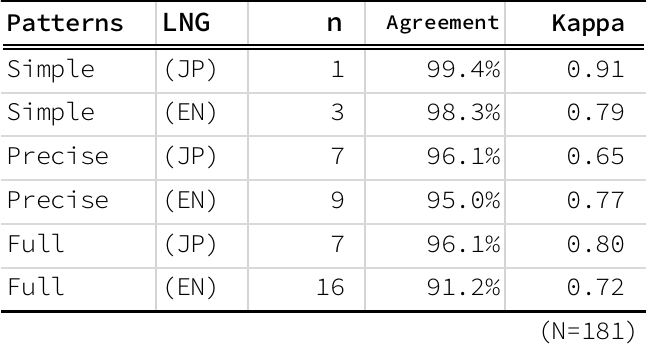}
\end{table}

\subsection{Predictive Performance}
First, we see the overall performance of annotation using ChatGPT. 
As mentioned above, ChatGPT was asked to annotate each prompt pattern twice, but here, we have taken the results of the first annotation. 
The accuracy of the results obtained by ChatGPT when given a prompt with the same content as the manual used in \citet{Nishikawa2023} was $61.3\%$ for citation purpose and $64.6\%$ for citation sentiment. 

Next, we look at the performance of the annotation results from the prompts with the highest consistency for each citation purpose and citation sentiment. 
For citation purpose, because the prompt that gave types of classes, their definitions, and annotation procedures in English (Precise, EN) were the most consistent, the relationship between the results of the first annotation with this pattern (Predict) and the gold standard (Actual) is summarized in \tabref{tab:pp_main}.

\begin{table}[htbp]
    \centering
    \caption{Predictive performance for citation purpose}
    \label{tab:pp_main}
    \includegraphics[width=80mm]{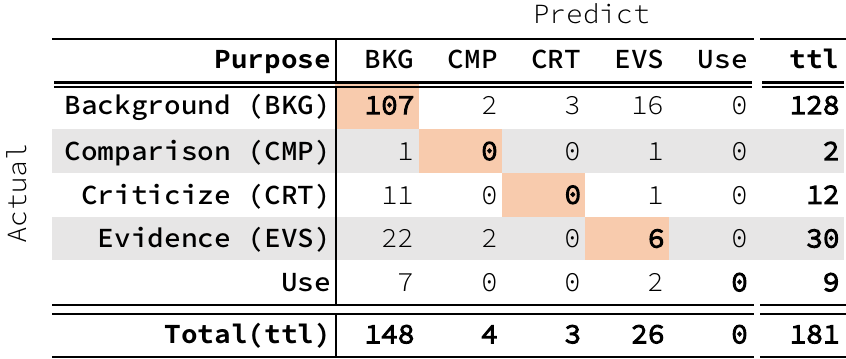}
\end{table}

\tabref{tab:pp_main} shows, for example, that when the correct answer is ``Background (BKG),'' ChatGPT correctly predicted BKG, i.e., True Positive, in 107 instances. 
However, a consistent number of errors are observed, including 16 cases where ChatGPT incorrectly predicted ``Evidence (EVS)'' instead of the correct BKG and 22 cases where ChatGPT predicted BKG, but the correct answer was EVS. 
Moreover, none of the ``Compare (CMP),'' ``Criticize (CRT),'' and ``Use'' were correctly predicted, even though they were originally infrequent classes.

Similarly, \tabref{tab:st_main} summarizes the relationship between the results of the first annotation and the prompt (Simple, EN), which provided only types of classes in English and was most consistent in the case of citation sentiment and the gold standard. 
The table shows, for example, that there are 120 cases where the correct answer is ``Neutral (NT)'' and it is correctly predicted as NT. 
However, there are some discrepancies, such as 24 cases where the correct answer was ``Positive (PG),'' even though it was predicted as NT.

\begin{table}[htbp]
    \centering
    \caption{Predictive performance for citation sentiment}
    \label{tab:st_main}
    \includegraphics[width=80mm]{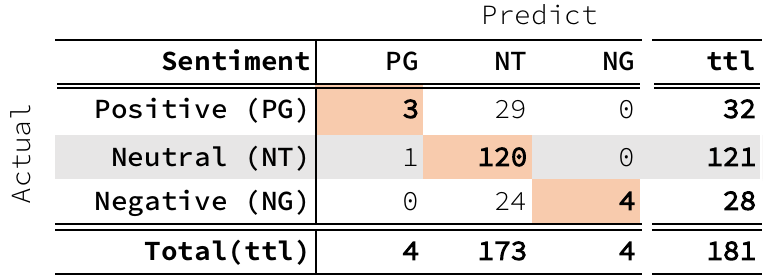}
\end{table}

Thus far, we have examined the results of the first annotation based on a single prompt. 
However, in the case of human annotation, it is common to create a single dataset for analysis, or in other words, a gold standard, by narrowing down the annotation results from multiple annotators through some means. 
Based on this, we had ChatGPT annotate using prompts of all types explained in the Methods section; then, we created a single dataset by integrating the results and compared the dataset with the gold standard. 
A majority vote was employed as the method of integration.

In \tabref{tab:pp_multi}, for citation purpose, the relationship between the ChatGPT dataset and the gold standard is summarized. 
Although there was a slight increase in the number of correct answers for the EVS in \tabref{tab:pp_multi}, the results were not significantly different from those shown in \tabref{tab:pp_main}. 

\begin{table}[htbp]
    \centering
    \caption{Predictive performance of ChatGPT's multiple annotations for citation purpose}
    \label{tab:pp_multi}
    \includegraphics[width=80mm]{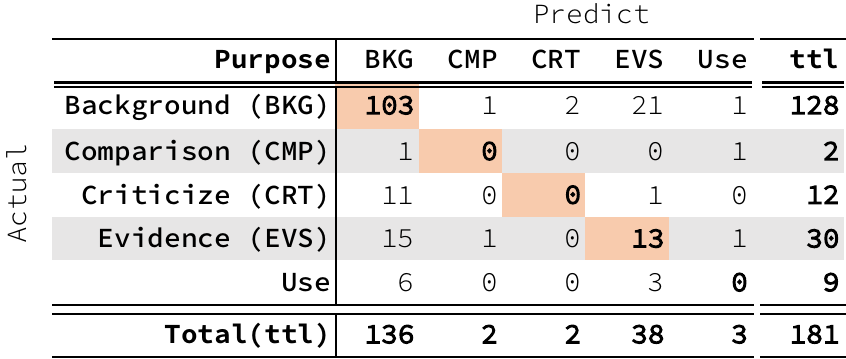}
\end{table}

Similarly, for citation sentiment, \tabref{tab:st_multi} shows the relationship between the results of ChatGPT’s annotation, merged into one by majority vote, and the gold standard. 
\tabref{tab:st_multi} shows that the performance for the prompt (Simple, EN) in \tabref{tab:st_main} and the performance when integrating the annotation results from all prompts are nearly comparable.

\begin{table}[htbp]
    \centering
    \caption{Predictive performance of ChatGPT's multiple annotations for citation sentiment}
    \label{tab:st_multi}
    \includegraphics[width=80mm]{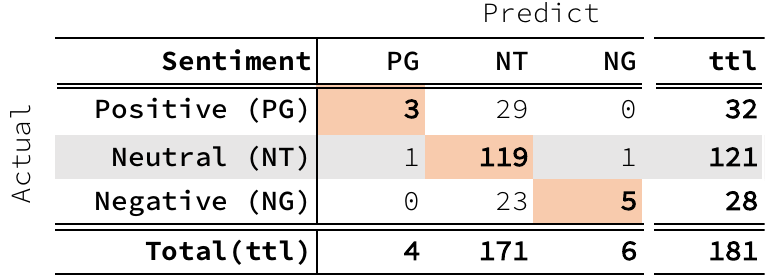}
\end{table}

\subsection{Discussion of Experimental Results}
The results of the experiments indicate that while ChatGPT outperforms human annotators in terms of consistency, it does not produce high-quality data in terms of predictive performance. 
Based on the results presented in \tabref{tab:pp_main} to \ref{tab:st_multi}, it can be said that ChatGPT does not predict the correct answer well, even though the gold standard created in \citet{Nishikawa2023} was originally highly skewed from class to class. 
For example, of the 148 cases in \tabref{tab:pp_main} predicted to be BKG, the number of cases that were actually BKG was 107 ($72.3\%$). 
Considering that the proportion of BKG in the correct data was $70.7\%$ as shown in \tabref{tab:pp_main}, the improvement rate is $1.6\%$ compared to the hypothetical case where ChatGPT always predicts any class as a BKG. 
In addition, it should be noted that there were a certain number of cases that were predicted as BKG but were not actually BKG, that is, false negatives, and those that were predicted as a class other than BKG but were actually BKG, that is false negatives.

As for citation sentiment, \tabref{tab:st_main} shows that all those predicted as PG are actually PG, which is a good prediction in terms of the small number of false negatives. 
However, this number is only 4 out of 28 total actual PG, which is not large. 
The same was true for NG. 
For NT, which accounts for the majority of cases, 120 of the 173 cases predicted as NT are actually NT, but the improvement rate is $1.5\%$ compared with the hypothetical case in which ChatGPT always predicts the class as NT. 
It should also be noted that approximately $30\%$ of the cases were falsely predicted to be PG or NG. 
Even if we consider the predicted results as PG or NG, this percentage is only approximately $4.4\%$ of the total. 
In other cases, the annotation results are unreliable; therefore, human review is inevitable.

Conversely, the results of the experiments in this study, which showed poor performance while maintaining a certain level of consistency, may indicate differences in how ChatGPT and human annotators ``interpret'' texts. 
Thus, we considered how ChatGPT interprets texts by examining the texts that were actually the target of the annotation in cases where ChatGPT failed to predict the classes correctly or where the results were inconsistent across multiple annotations.

The results of the examination suggest that there is no explicit expression of the relationship between the target cited paper and its surrounding sentences in the texts for which ChatGPT makes erroneous or inconsistent annotations. 
For example, ChatGPT predicted the class as ``Criticism (CRT)'' for the following text \citep[][p.207]{Borie2019}\footnote{It is the citaion pair (NS-SSH) in SDG13.}, but the correct class was "Background (BKG)".

\begin{quote}
    (...)
    This use of technical devices in an attempt to suppress political debates has been widely documented elsewhere (e.g. Latour, 2004; Lupton and Mather, 1997). 
    At an extreme this use of GIS and mapping reinforces and aggravates existing divides and inequalities.
\end{quote}

In this text, the target-cited paper was \citet{Lupton1997}. 
Although the text includes words that would provide a basis for predicting the class as CRT (``widespread'' and ``reinforces and exacerbates existing divisions and inequalities''), it is ``this use of technical devices/GIS and mapping,'' not \citet{Lupton1997}, that is the target of ``criticism'' in this text. 
Because \citet{Lupton1997} seems to have been cited to provide background information on the research topic of the citing paper, the correct class here is the BKG. 
This would have been relatively easy for human annotators to determine, but it would have been difficult for ChatGPT to do so because the relationship between the cited and citing papers was not explicitly stated as words. 
In other words, it is suggested that ChatGPT interprets text using only explicit words and does not consider implied contexts. 
Note that this pattern is often seen in texts where ChatGPT fails to correctly predict the class, but there are exceptions in which this pattern does not successfully explain the reason for the interpretation.

From the above, it can be said that the annotation results of ChatGPT are inadequate in terms of performance, and it is problematic to use the dataset created by ChatGPT for analysis. 
In other words, the experimental results of this study clarify that it is difficult to use the current ChatGPT as a substitute for human annotators in citation context analyses.

\section{Consideration of LLM Use Case in Citation Context Analysis}

\subsection{Support for Human Annotators}
Thus far, we have evaluated the LLM from the viewpoint of consistency and predictive performance when instructions are given based on a manual for humans. 
Consequently, the annotation results of the LLM are not likely to be as good as those of humans in terms of performance, and it was suggested that human annotation is necessary for unknown data.

However, it may be possible to use LLMs not as a complete replacement for human annotation, but as a support for human annotators. 
If we can believe that ``what LLMs predict as PG is actually PG,'' as in the case of PG in \tabref{tab:pp_main}---i.e., if the prediction performance for at least some classes is sufficiently high---we can let LLMs annotate them on behalf of human annotators. 
Moreover, as with human annotations, it is also possible that some of those predicted to be in the same class include those predicted with confidence, whereas others do not. 
In this case, it may be possible to reduce the cost of annotation by having low-confidence prediction results annotated again by a human and adopting LLMs’ annotation results of the LLMs for high-confidence prediction results.

In general, the number of possible classes affects the text classification performance, and the annotation performance by LLMs depends on the type of class predicted \citep{Pangakis2023}. 
Therefore, the performance of LLMs annotation can be improved by changing the number of classes and their characteristics. 
In light of the above, we reviewed prompts for using LLMs to support human annotators and examined the possibility of their use.

\subsubsection{Citation Purpose}
``Background (BKG)'' accounts for about $70\%$ of the gold standard, followed by ``Evidence'' at about $16\%$. 
We thus reconfigured the annotation for citation purpose as a three-class classification task, adding ``Other'' to these two classes. 
\tabref{tab:pp_c_01} presents the annotation results for ChatGPT\footnote{The prompts used for the following tasks in this section are shown in Online Resource 2.}.

\begin{table}[htbp]
    \centering
    \caption{Predictive Performance of ChatGPT's Annotation for Citation Purpose when the Third Class is "Other"}
    \label{tab:pp_c_01}
    \includegraphics[width=80mm]{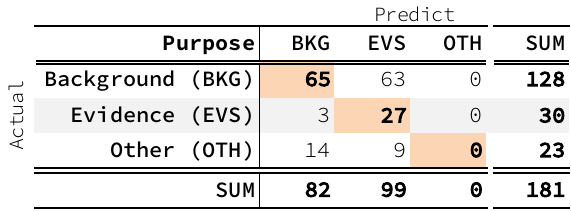}
\end{table}

As shown in the table, none of the cases were classified as ``Other,'' and the predicted results are either ``Background'' or ``Evidence.''
Compared to \tabref{tab:pp_main} and \tabref{tab:pp_multi}, the number of cases predicted as ``Evidence'' increased, and the performance of annotation generally worsened. 
In addition, the consistency (simple agreement rate) of the two annotations is $91.7\%$. 
One possible reason for this change in annotation trends is that ChatGPT was influenced by the literal meaning of the name of the newly created class ``Other'' and avoided annotating that broad and ambiguous class. 
We thus replaced ``Other'' with ``General'' or ``Pending,'' respectively, and let ChatGPT annotate again.

As shown in \tabref{tab:pp_c_02} and \ref{tab:pp_c_03}, although some cases belonging to the third class can be observed, the overall trend is the same as that shown in \tabref{tab:pp_c_01}. 
In addition, the consistency (simple agreement rate) of the results of the two annotations was $86.2\%$ and $91.7\%$.

\begin{table}[htbp]
    \centering
    \caption{Predictive Performance of ChatGPT's Annotation for Citation Purpose when the Third Class is "General"}
    \label{tab:pp_c_02}
    \includegraphics[width=80mm]{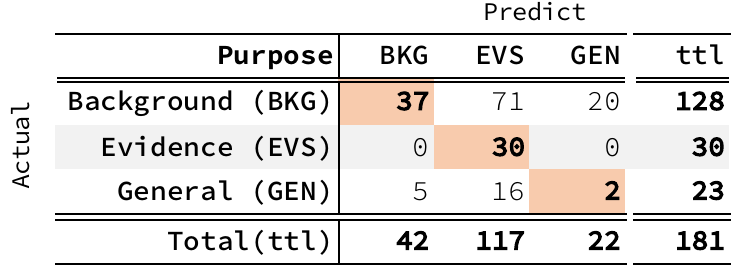}
\end{table}

\begin{table}[htbp]
    \centering
    \caption{Predictive Performance of ChatGPT's Annotation for Citation Purpose when the Third Class is "Pending"}
    \label{tab:pp_c_03}
    \includegraphics[width=80mm]{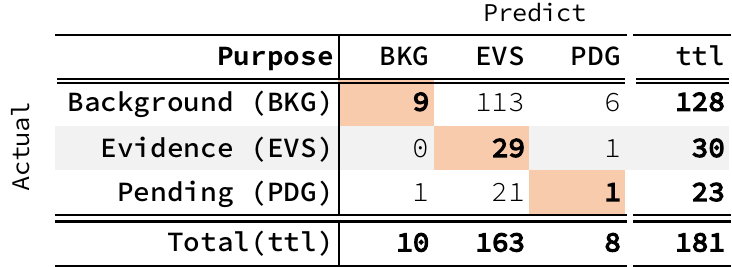}
\end{table}

The above experiments were conducted hoping that reducing the number of classes would improve the annotation performance; however, the results showed that the performance degraded. 
We also changed the third-class names to account for the possibility that the name might affect annotation; however, performance did not improve. 
However, if the literal meaning of the class name affects the annotation, it is possible that the existing classes ``Background'' and ``Evidence'' also affected the annotation trend of ChatGPT, apart from their operative definition. 
Therefore, we let ChatGPT annotate again by replacing ``Background'' and ``Evidence'' with just ``BKG'' and ``EVS,'' which are sequences of symbols without meaning as words. 
Moreover, the name of the third class was changed to UKN.
The annotation results for these changes are listed in \tabref{tab:pp_c_04}. 
The overall trend was the same as that shown in \tabref{tab:pp_c_01}. 
In addition, the consistency (simple agreement rate) of the results for the two annotations is $86.2\%$.

\begin{table}[htbp]
    \centering
    \caption{Predictive Performance of ChatGPT's Annotation for Citation Purpose with Meaningless Class Names}
    \label{tab:pp_c_04}
    \includegraphics[width=80mm]{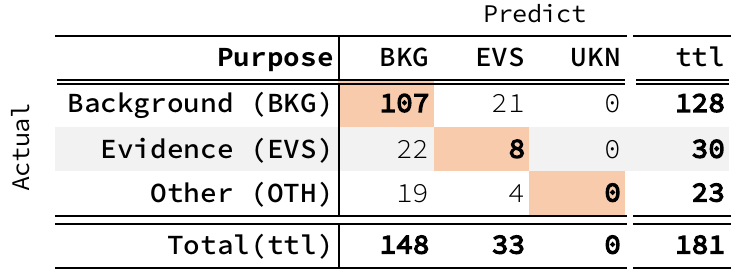}
\end{table}

The experiments thus far have shown that reducing the number of classes or changing class names does not improve the annotation performance. 
Therefore, we attempted a different strategy, having GPT perform a binary classification for each class. 
Specifically, we let ChatGPT predict whether it was a BKG (PB or NB) or an EVS (PE or NE) and examined the relationship between their combination and the gold standard. 
Although we replaced the original annotation with a binary classification, there was no relationship, as shown in \tabref{tab:pp_c_05}. 
In addition, the consistency (simple agreement rate) of the results of the two annotations was $90.6\%$ and $91.7\%$.

\begin{table}[htbp]
    \centering
    \caption{Predictive Performance of ChatGPT's Annotation for Citation Purpose when Binary Classification for Each Class}
    \label{tab:pp_c_05}
    \includegraphics[width=92mm]{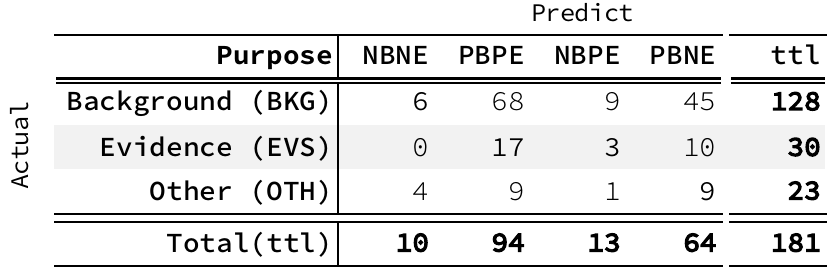}
\end{table}

Finally, we reorganized the annotation into the simplest binary classification: BKG (BKG or UKN). 
\tabref{tab:pp_c_06} shows that of the 43 cases predicted as BKG, 38 ($88.4\%$) were BKG, which is highly accurate. 
Although the number of cases in which ChatGPT answered correctly was small, these data could potentially be used in the analysis. 
In this case, the consistency (simple agreement rate) between the two annotations is $91.2\%$.

\begin{table}[htbp]
    \centering
    \caption{Predictive Performance of ChatGPT's Annotation for Citation Purpose when Binary Classification}
    \label{tab:pp_c_06}
    \includegraphics[width=68mm]{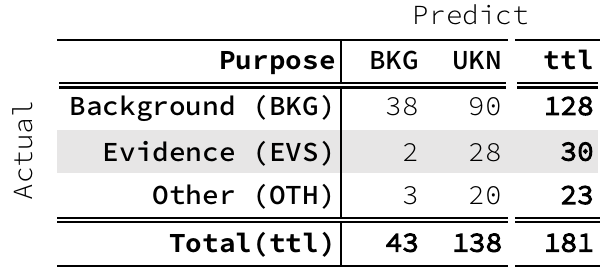}
\end{table}

\subsubsection{Citation Sentiment}
Unlike citation purpose, citation sentiment was originally a three-class classification task, and it was difficult to reduce the number of classes any further. 
We thus attempted a different approach: adding a class. 

As mentioned previously, some of the predicted results from the LLMs may contain different confidence levels. 
If so, the annotation performance could be improved by distinguishing between those predicted with high and low confidence. 
Therefore, we first added a class named ``Pending (PD)'' and let those with low confidence be classified there.
As shown in \tabref{tab:st_c_02}, some are classified as ``Pending,'' but the number is small, 30. 
Furthermore, what is actually a ``Positive'' may be classified as the opposite, ``Negative,'' which means that the performance is deteriorating.

\begin{table}[htbp]
    \centering
    \caption{Predictive Performance of ChatGPT's Annotation for Citation Sentiment when Adding "Pending"}
    \label{tab:st_c_02}
    \includegraphics[width=82mm]{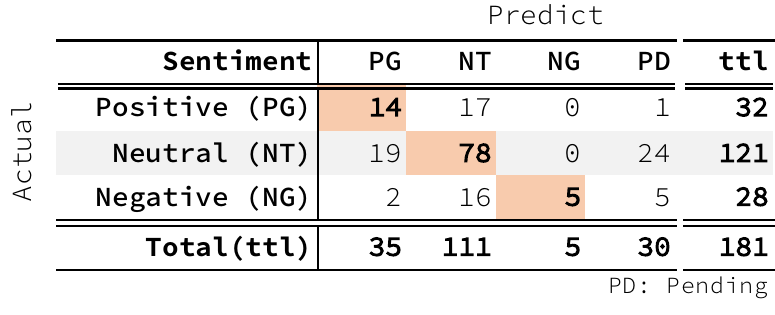}
\end{table}

Next, we had both those that were actually neutral and those predicted with low confidence be classified in the current ``Neutral,'' and we have renamed only this class to UKN, a name without meaning. 
As shown in \tabref{tab:st_c_03}, while the number of those predicted as ``Positive'' and ``Negative'' has increased, the number of errors was too large to use the results for analysis.

\begin{table}[htbp]
    \centering
    \caption{Predictive Performance of ChatGPT's Annotation for Citation Sentiment when Changing "Neutral"}
    \label{tab:st_c_03}
    \includegraphics[width=80mm]{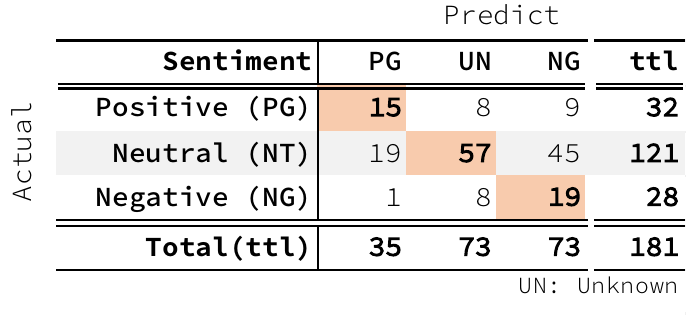}
\end{table}

\subsection{Other Use Cases}
The results of the experiments thus far indicate that it is difficult to use LLMs to perform annotations partially on behalf of humans. 
However, they also suggest that there is room for LLMs to be utilized in citation context analysis.

\citet{Pangakis2023} categorizes the use cases of LLMs in general annotation work as follows:

\begin{enumerate}
    \item Confirming the quality of human-labeled data
    \item Identifying cases to prioritize for human review
    \item Producing labeled data to finetune and validate a supervised classifier
    \item Classifying the entire corpus directly
\end{enumerate}
This paper shows that cases other than Case 1 are difficult to apply to citation context analysis. 
For Case 4, as we have seen in the Experiments section, poor predictive performance makes it problematic to use the LLM-generated data for the analysis. 
For the same reason, the use of LLMs in Case 3 should be avoided. 
In addition, Case 2 is a use case related to the partial substitution of annotation work, but as discussed thus far in this section, there are also concerns about the use of LLMs for this purpose.

However, Case 1 seems open for consideration. 
This use case implies examining the quality of human annotation results by comparing the annotation results of humans and the LLM. 
Putting this into the context of citation context analysis, a possible use of LLMs is to use the results of the LLMs annotation as reference information when narrowing down the data produced by multiple human annotators to a single set of data to be used in the analysis. 
As shown in the Experiments section, although the performance of the annotation results by the LLM was low, the consistency was more stable than that of the human annotators. 
The LLM can also output reasons for its decisions. 
Because of these characteristics, the LLM can be viewed as an annotator with criteria and tendencies different from those of human annotators. 

In other words, the annotation results of LLMs and the reasons for their decisions can be utilized as reference information in the process of the aforementioned ``discussion''\citep{Lin2018}, which narrows down multiple datasets to a single one. 
In ``discussion,'' human annotators try to maintain the objective correctness of the data by explaining the reasons for their decisions of annotations to each other, taking care not to persuade the other, and voluntarily revising their own work results if necessary. 
At this time, using LLMs data as reference information from a third-party standpoint may reduce the possibility of a particular human annotator’s subjectivity having a significant impact on the ``discussion.''

Additionally, hiring human annotators is generally expensive in terms of time and money, which sometimes forces the use of data from a single annotator for analysis \citep{Rytting2023}. 
Particularly in the case of citation context analysis, annotators must have more advanced skills because the text to be annotated is a special one, a scientific paper. 
This makes it more challenging to secure a sufficient number of annotators compared with general annotations. 
In these situations, one option would be to introduce LLMs as one of the annotators to avoid using the data generated by a single annotator.

\section{Conclusion}
This study aimed to explore the applicability of LLMs to citation context analysis. 
The results revealed that ChatGPT, at least in its current version, cannot annotate with sufficiently high performance to replace human annotators for the major categories in citation context analysis: citation purpose and sentiment. 
It was also found to be difficult to have ChatGPT partially annotated on behalf of humans, such as by using CatGPT annotation results for specific classes and having a human annotate the rest.

However, because the ChatGPT annotation results have a certain consistency, it may be possible to view LLMs as annotators who interpret differently than humans. 
This suggests the following two possible use cases of LLMs in citation context analysis. 
First, the annotation results obtained by LLMs can be used as reference information when narrowing the annotation results obtained by multiple human annotators to one. 
Second, it is possible to use LLMs as the Nth annotators when securing the number of human annotators is difficult.

Future researchers attempting to utilize LLMs for citation context analysis can refer to LLMs' limitations and use cases clarified in this study. 
In contrast to previous studies that verified the performance of LLMs in general annotation tasks and proposed use cases, the findings of this study are novel in that they examined their applicability and use cases in the specific task of citation context analysis. 

However, this study had several limitations. 
In the present study, we focused on \texttt{gpt3-turbo-0310}, which exhibited the best performance at the time of the experiment. 
However, \texttt{gpt4}, which is believed to perform better than \texttt{gpt3-turbo-0310}, is now available to the public. 
We also tried an experiment using this new model and found that it is worse than \texttt{gpt3-turbo-0310} in terms of predictive performance, as shown in Online Resource 3; however, further new models may emerge in the future that will allow LLMs to perform in a way that overturns the conclusions of this study. 
Therefore, the findings of this study represent a snapshot of the potential applications of LLMs and should be analyzed continuously following future technological trends.

\clearpage
\section*{Acknowledgements}
This preprint has not undergone peer review (when applicable) or any post-submission improvements or corrections.
The Version of Record of this article is published in Scientometrics, and is available online at \url{https://doi.org/10.1007/s11192-024-05142-9}

\section*{Declarations}
\subsection*{Competing Interests} 
The authors have no competing interests to declare that are relevant to the content of this article.

\subsection*{Funding} 
No funding was received for conducting this study.

\clearpage
\bibliographystyle{plainnat}
\bibliography{ref}

\appendix

\clearpage
 \section{Online Resource 1. Prompts used in Experiments}

\noindent
\textbf{Note:}
\begin{quotation}
The following shows only the prompts given to ChatGPT that were written in English.
In each prompt, the actual data to be annotated is inserted in the area enclosed in \{ \}.
\end{quotation}

\subsection{Citation Purpose}

\subsubsection{Simple}
\begin{quote}
Classify the type of purpose for which the author of the following text is citing the target.

The type of citation purpose category is Background, Comparison, Criticize, Evidence, or Use.

The label of the target literature is Target: onwards, up to a new line.
The thesis is the text after Text:.

Just report the classification result.

Target: \{ \} \\
Text:\\
 \{ \} 
\end{quote}

\subsubsection{Basic}
\begin{quote}
Classify the type of purpose for which the author of the following text is citing the target.

The type of citation purpose category is Background, Comparison, Criticize, Evidence, or Use based on the following manual.

Manual: \\
----------- \\
If the target is cited to present or summarize general background information about the research theme or topic of the text, you classify the text as Background.

If the target is cited to compare results or methods between the text and the target or between other cited papers, you classify the text as Comparison.

If the target is cited to provide some evaluation or review of the text, you classify the text as Criticize. Both positive and negative evaluations are included here.

If the target is cited to support or validate the author's claims, decisions (e.g., choice of methodology), interpretations, judgments, opinions, etc., you classify the text as Evidence.

If the target is cited to use methods, models, data, software, concepts, theories, hypotheses, etc. presented in the target, you classify the text as Use.\\
-----------

The label of the target literature is Target: onwards, up to a new line.

The thesis is the text after Text:.
Just report the classification result.

Target: \{ \} \\
Text:\\
 \{ \} 
\end{quote}

\subsubsection{Precise}
\begin{quote}
Classify the type of purpose for which the author of the following text is citing the target based on the following manual.

The type of citation purpose category is Background, Comparison, Criticize, Evidence, or Use.

Manual: \\
-----------\\ 
The classification procedure is as follows: 

1. Read the sentence containing the target first. If the category can be clearly determined by reading the sentence containing the target, the category is determined at that point.

2. If you are not sure about the decision based on the sentence containing the target alone, read the one sentence before and after it. If the category can be clearly determined by reading the sentences before and after the sentence containing the target, the category is determined at that point.

3. If the category cannot be determined after reading one sentence before or after the sentence containing the target, read the sentences in order from the beginning of the paragraph. If the category can be clearly determined, the category is determined at that point.
If the target is cited to present or summarize general background information about the research theme or topic of the text, you classify the text as Background.

If the target is cited to compare results or methods between the text and the target or between other cited papers, you classify the text as Comparison.

If the target is cited to provide some evaluation or review of the text, you classify the text as Criticize. Both positive and negative evaluations are included here.

If the target is cited to support or validate the author's claims, decisions (e.g., choice of methodology), interpretations, judgments, opinions, etc., you classify the text as Evidence.

If the target is cited to use methods, models, data, software, concepts, theories, hypotheses, etc. presented in the target, you classify the text as Use.\\
-----------

The label of the target literature is Target: onwards, up to a new line.

The thesis is the text after Text:.
Just report the classification result.

Target: \{ \} \\
Text:\\
 \{ \} 
\end{quote}

\subsubsection{Full}

\begin{quote}
Classify the type of purpose for which the author of the following text is citing the target based on the following manual.

The type of citation purpose category is Background, Comparison, Criticize, Evidence, or Use.

I will give you the text later.

Manual: \\
----------- \\
The classification procedure is as follows: 

1. Read the sentence containing the target first. If the category can be clearly determined by reading the sentence containing the target, the category is determined at that point.

2. If you are not sure about the decision based on the sentence containing the target alone, read the one sentence before and after it. If the category can be clearly determined by reading the sentences before and after the sentence containing the target, the category is determined at that point.

3. If the category cannot be determined after reading one sentence before or after the sentence containing the target, read the sentences in order from the beginning of the paragraph. If the category can be clearly determined, the category is determined at that point.

If the target is cited to present or summarize general background information about the research theme or topic of the text, you classify the text as Background.

If the sentences around the target contain the following keywords, the text is considered Background.

Keywords of Background: ``overview'', ``review'', ``summarize''

The following are the criteria for classifying the text as Background.

Criteria of Background: ``When the target is cited to summarize information on recent research trends'', ``If the text simply introduces or refer to the target on its research topic'', ``When the target is cited to state general or overall information (e.g., policy trends, relevant research areas or theories, etc.)'', ``If the text does not fall into any other category''

If the target is cited to compare results or methods between the text and the target or between other cited papers, you classify the text as Comparison.

If the sentences around the target contain the following keywords, the text is considered Comparison.

Keywords of Comparison: ``although'', ``compare'', ``comparison'', ``contrast'', ``however'', ``in contrast'', ``on the contrary'', ``on the other hand'', ``while''

The following are the criteria for classifying the text as Comparison.

Criteria of Comparison: ``When cited to compare the results of the citing paper with those of previous studies and claim the superiority of the citing paper's results'', ``When comparing two previous studies and pointing out the advantages or disadvantages of one study over the other''

If the sentences around the target are similar to the following examples, the text is considered Comparison.

An example of Comparison: ``However, while neurobiology posits that the rewarding properties of social behavior may have evolved to facilitate group cohesion and cooperation [4], our model suggests that polarization (as opposed to cohesion) across groups may be a side-effect of these rewarding properties.''

If the target is cited to provide some evaluation or review of the text, you classify the text as Criticize. Both positive and negative evaluations are included here.

The following are the criteria for classifying the text as Criticize.

Criteria of Criticize: ``When the target is cited to evaluate the contribution or advantage of the text'', ``When the text is cited to point out a weakness or wrong in the text''

If the sentences around the target are similar to the following examples, the text is considered Criticize.

An example of Criticize: ``The method in [4] reports a high result for the Media-lab dataset but does this using a dataset-specific SE and so it not a universal method.''

If the target is cited to support or validate the author's claims, decisions (e.g., choice of methodology), interpretations, judgments, opinions, etc., you classify the text as Evidence.

If the sentences around the target contain the following keywords, the text is considered Evidence.

Keywords of Evidence: ``aligns with'', ``be consistent with'', ``indicate to us'', ``similar to'', ``support'', ``therefore'', ``thus''

The following are the criteria for classifying the text as Evidence.

Criteria of Evidence: ``When the target is cited to support the text’s author’s claim, hypothesis, or decision'', ``When the target is cited to justify the methodology of the text’s research or previous research the text’s author supports'', ``When the target is cited to justify the assumptions and limitations of the text'', ``When the text’s author proposes a future research direction, the author cites the target to support his proposal''

If the sentences around the target are similar to the following examples, the text is considered Evidence.

An example of Evidence: ``Our findings emphasize that building digital seizing capabilities are contingent on pacing strategic actions, which aligns with dynamic capabilities research in hypercompetitive contexts [4].''

If the target is cited to use methods, models, data, software, concepts, theories, hypotheses, etc. presented in the target, you classify the text as Use.

If the sentences around the target contain the following keywords, the text is considered Use.

Keywords of Use: ``based on'', ``be carried over'', ``provided by'', ``use''

The following are the criteria for classifying the text as Use.

Criteria of Use: ``When the target is cited to use a dataset presented in the target'', ``When the target is cited to use a method proposed or developed in the target'', ``When the author of the text cites a definition on a concept or theory presented in the target''

If the sentences around the target are similar to the following examples, the text is considered Use.

An example of Use: ``Our Arabic part-of-speech tagger uses the simplified PATB tag set proposed by [4].''\\
-----------

The label of the target literature is Target: onwards, up to a new line.

The thesis is the text after Text:.
Just report the classification result.

Target: \{ \} \\
Text:\\
 \{ \} 
\end{quote}

\subsection{Citation Sentiment}

\subsubsection{Simple}
\begin{quote}
Classify the text into Positive, Negative, or Neutral.

The label of the target literature is Target: onwards, up to a new line.

The thesis is the text after Text:.
Just report the classification result.

Target: \{ \} \\
Text:\\
 \{ \} 
\end{quote}

\subsubsection{Precise}
\begin{quote}
Classify the text into Positive, Negative, or Neutral based on the following manual. 

Manual:\\
-----\\
You classify the sentence in which the target is cited. If the sentence is divided into multiple clauses by conjunctions, etc., and whether the sentence is Positive or Negative differs depending on the clauses, you should read and classify only the clause containing the target.

If the target is cited in a sentence with a positive meaning, you classify it as Positive.

If the target is cited in a sentence with a negative meaning, you classify it as Negative.

If the target is cited in a sentence with neither Positive nor Negative meaning, you classify it as Neutral.\\
-----

The label of the target literature is Target: onwards, up to a new line.
The thesis is the text after Text:.
Just report the classification result.

Target: \{ \} \\
Text:\\
 \{ \} 
\end{quote}

\subsubsection{Full}
\begin{quote}
Classify the text into Positive, Negative, or Neutral based on the following manual. 

Manual:\\
-----\\
You classify the sentence in which the target is cited. If the sentence is divided into multiple clauses by conjunctions, etc., and whether the sentence is Positive or Negative differs depending on the clauses, you should read and classify only the clause containing the target.

If the target is cited in a sentence with a positive meaning, you classify the text as Positive.

Sentences containing the following keywords are considered Positive.

Examples of Positive keywords: ``be able to...'', ``best'', ``can...'', ``could'', ``develop'', ``enhance'', ``important'', ``promote'', ``robustly'', ``support'', ``well''

The following sentences are examples of sentences that are considered Positive.

Example of Positive sentences: ``The best known and simplest stochastic representation for discrete geophysical time series is the AR(1) model (Ghil et al. 2002; Bretherton and Battisti 2000).''

Example of Positive sentences: ``These patterns find empirical support in Popp and Newell’s (2012) study of firm-level R\&D spending and patents.''

Example of Positive sentences: ``Although national and transnational connections may be necessary to secure access to resources and technical expertise, it is argued that local participation in the governance of social-ecological systems provides legitimacy (Biermann and Gupta 2011, Dryzek and Stevenson 2011), accommodates diverse interests and values (Brown 2003, Lebel et al. 2006), and taps local ecological knowledge (Berkes and Folke 2002, Gerhardinger et al. 2009, Raymond et al. 2010).''

If the target is cited in a sentence with a negative meaning, you classify the text as Negative.

Sentences containing the following keywords are considered Negative.

Examples of Negative keywords: ``but'', ``despite'', ``even though'', ``however'', ``ignore'', ``less'', ``nevertheless'', ``problematic'', ``suffer'', ``undermine''

The following sentences are examples of sentences that are considered Negative.

Example of Negative sentences: ``When women are unable to obtain sufficient water for menstrual ablutions or hygiene (e.g., cleaning menstrual cloths), they may suffer extreme stigma and humiliation (Rashid and Michaud 2000:54).''

Example of Negative sentences: ``The need for better empirical information about energy-efficiency R\&D is well known but difficult to solve due to lack of disaggregated data (although see on the contrary Popp (2002) and Popp and Newell (2012)).''

Example of Negative sentences: ``Determination of the functions of fungal species, which typically requires their isolation in pure culture and the study of their effects on defined substrates, has well-documented limitations [19–21].''

If the target is cited in a sentence with neither Positive nor Negative meaning, you classify the text as Neutral.\\
-----

The label of the target literature is Target: onwards, up to a new line.
The thesis is the text after Text:.
Just report the classification result.

Target: \{ \} \\
Text:\\
 \{ \} 
\end{quote}

\clearpage
\section{Online Resource 2. Prompts used in Consideration of LLM Use Case in Citation Context Analysis}

\noindent
\textbf{Note:}
\begin{quotation}
Below is each prompt used in Consideration of LLM Use Case in Citation Context Analysis.
Each prompt is named with the corresponding table number in the article.
In each prompt, the actual data to be annotated is inserted in the area enclosed in \{ \}.
\end{quotation}

\subsection{Citation Purpose}

\subsubsection{Table 13}
\begin{quote}
Please classify the type of purpose for which the author of the following text is citing the target. 

The type of citation purpose category is Background, Evidence, or Others based on the following manual. 

Manual:  \\
--- \\
If the target is cited to present or summarize general background information about the research theme or topic of the text, you classify the text as Background. 

If the target is cited to support or validate the author's claims, decisions (e.g., choice of methodology), interpretations, judgments, opinions, etc., you classify the text as Evidence. 

If the type of citation purpose category is neither Background nor Evidence, you classify the text as Others. \\
--- 

The label of the target literature is Target: onwards, up to a new line.

The thesis is the text after Text:.
Just report the classification result.

Target: \{ \} \\
Text:\\
 \{ \} 
\end{quote}

\subsubsection{Table 14}
\begin{quote}
Please classify the type of purpose for which the author of the following text is citing the target. 

The type of citation purpose category is General, Background or Evidence based on the following manual. 
 
Manual:  \\
--- \\
Usually annotate as 'General'. 

For example,  
1. a comparison of results or methods between the text and the subject or other cited papers. 

2. some kind of evaluation or review of the text. 

3. use of methods, models, data, software, concepts, theories, hypotheses, etc. 

All other items that are not background, evidence, etc., that cannot definitely be said to be background or evidence, or that cannot be classified in one specific category, etc., all belong to the 'General' category. 

In addition, 
If the target is cited to present or summarize general background information about the research theme or topic of the text, you classify the text as Background. 

If the target is cited to support or validate the author's claims, decisions (e.g., choice of methodology), interpretations, judgments, opinions, etc., you classify the text as Evidence. \\
--- 

The label of the target literature is Target: onwards, up to a new line.

The thesis is the text after Text:.
Please, return only the classification results in one word.

Target: \{ \} \\
Text:\\
 \{ \} 
\end{quote}

\subsubsection{Table 15}
\begin{quote}
Please classify the type of purpose for which the author of the following text is citing the target. 

The type of citation purpose category is Pending, Background or Evidence based on the following manual. 
 
Manual:  \\
--- \\
Wherever possible, judge the case as 'Pending'. 

Anything that cannot definitely be categorised as Background or Evidence, or that cannot be placed in a specific category, belongs to 'Pending'. 
In addition, 

If the target is cited to present or summarize general background information about the research theme or topic of the text, you classify the text as Background. 

If the target is cited to support or validate the author's claims, decisions (e.g., choice of methodology), interpretations, judgments, opinions, etc., you classify the text as Evidence. \\
--- 

The label of the target literature is Target: onwards, up to a new line.
The thesis is the text after Text:.
Please, return only the classification results in one word.

Target: \{ \} \\
Text:\\
 \{ \} 
\end{quote}

\subsubsection{Table 16}
\begin{quote}
Please provide a classification of the citation in the text.

Classification target is only one citation.

The label of the target citation is [Target:] onwards, up to a new line.
The text is after [Text:].

The criteria for classification are as follows.

There are three classes: UKN, BKG and EVS.

The criteria for classification are given below.

Basically, classified as UKN.

This option (i.e. UKN) is extremely strongly recommended.

Try to select this option whenever possible.

Where the purpose of the citation is presumed to be background, such as including present or summarize general background information about the research theme or topic, 
it is classified as BKG.

This option (i.e. BKG) is second most strongly recommended.

Where the purpose of the citation is presumed to be evidence, such as support or validate the author's claims, decisions (e.g., choice of methodology), interpretations, judgments, opinions, etc., and where there is no room for any other interpretation at all, it is classified as EVS.

This option (i.e. EVS) is deprecated, avoid as much as possible.

Please, return only the classification results in just one word.

Target: \{ \} \\
Text:\\
 \{ \} 
\end{quote}

\subsubsection{Table 17}
\begin{quote}
Please provide a classification of the citation in the text.

Classification target is only one citation.

The label of the target citation is [Target:] onwards, up to a new line.
The text is after [Text:].

The criteria for classification are as follows.

There are two types of classes.

Basically, classified as NB.

However, where the purpose of the citation is presumed to be background, such as including present or summarize general background information about the research theme or topic, 
it is classified as PB.
 
Please, return only the classification results in just one word.

Target: \{ \} \\
Text:\\
 \{ \} 
\end{quote}

\subsubsection{Table 18}
\begin{quote}
Please provide a classification of the citation in the text.

Classification target is only one citation.

The label of the target citation is [Target:] onwards, up to a new line.
The text is after [Text:].

The criteria for classification are as follows.

There are 2 classes: UKN and BKG.

The criteria for classification are given below.

Basically, classified as UKN.

This category includes, for example, citations as evidence, criticism, comparison, discussion, etc.

Try to select this option whenever possible.

The other hands, Where the purpose of the citation is presumed to be background, such as including present or summarize general background information about the research theme or topic, it is classified as BKG.

Please, return only the classification results in just one word.

Target: \{ \} \\
Text:\\
 \{ \} 
\end{quote}

\subsection{Citation Sentiment}

\subsubsection{Table 19}
\begin{quote}
Please classify the text into Positive, Negative, Neutral or Pending based on the following manual.  
 
Manual: \\
---- \\
You classify the sentence in which the target is cited.  

Wherever possible, judge the case as 'Pending'. 

If the target is cited in a sentence with a definitely positive meaning, you classify it as 'Positive'. 

If the target is cited in a sentence with a definitely negative meaning, you classify it as 'Negative'. 

If the target is cited in a sentence with a definitely neutral meaning, you classify it as 'Neutral'. \\
---- 

The label of the target literature is Target: onwards, up to a new line.

The thesis is the text after Text:.
Just report the classification result.

Target: \{ \} \\
Text:\\
 \{ \} 
\end{quote}

\subsubsection{Table 20}
\begin{quote}
Please provide a classification of the citation in the text.

Classification target is only one citation.

The label of the target citation is [Target:] onwards, up to a new line.
The text is after [Text:].

The criteria for classification are as follows.

There are 3 classes: Positive, Negative and UKN.

The criteria for classification are given below.

Basically, classify them as UKN.

This category also includes the classification result of neutral.

However, if the target citation is presumed to be cited positively or negatively, output the presumption.

Please, return only the classification results in just one word.

Target: \{ \} \\
Text:\\
 \{ \} 
\end{quote}

\clearpage
\section{Online Resource 3. Examples of Annotations with GPT4}

\noindent
\textbf{Note:}
\begin{quotation}
At the start of the experiment (May 2023), of the models available to the general public via the API, the most capable model was \texttt{gpt-3.5-turbo} and was therefore used in this part.

However, GPT4 has also been available since July 2023.
As GPT4 is generally considered to have higher performance than \texttt{gpt-3.5-turbo}, in the following, we show the results of having GPT4 annotated.

Specifically, we had GPT4 annotated in the same way as discussed in Experiments.
However, only English prompts were used.
\end{quotation}

\subsection{Citation purposes}
The annotation results for citation purpose are summarized in \tabref{tab:gpt4_p1} through \tabref{tab:gpt4_p4}.

\begin{table}[htbp]
    \centering
    \caption{Citation purposes: Simple (English)}
    \label{tab:gpt4_p1}
    \includegraphics[width=80mm]{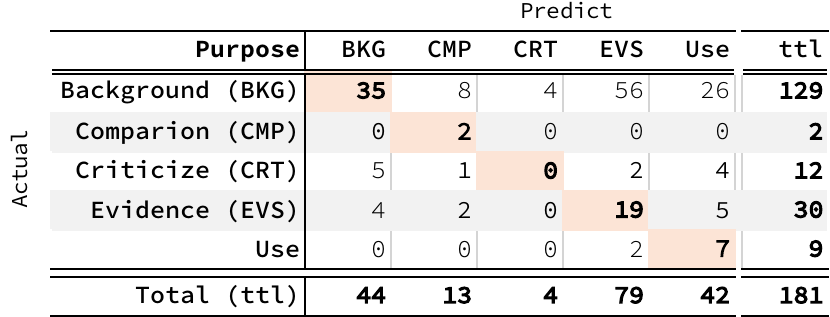}
\end{table}

\begin{table}[htbp]
    \centering
    \caption{Citation purposes: Basic (English)}
    \label{tab:gpt4_p2}
    \includegraphics[width=80mm]{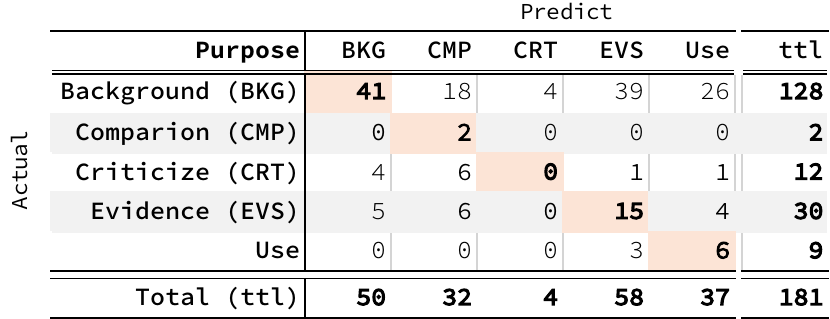}
\end{table}

\begin{table}[htbp]
    \centering
    \caption{Citation purposes: Precise (English)}
    \label{tab:gpt4_p3}
    \includegraphics[width=80mm]{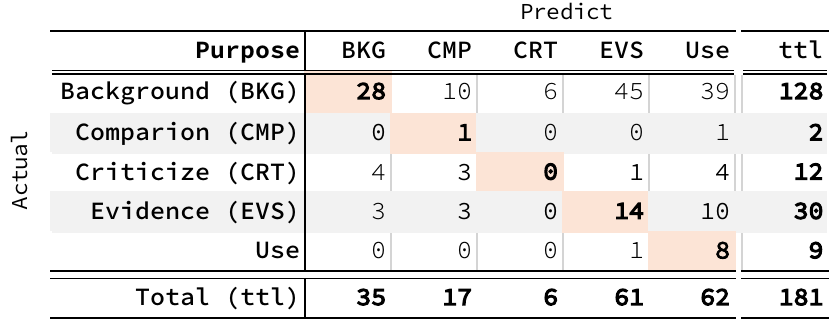}
\end{table}

\begin{table}[htbp]
    \centering
    \caption{Citation purposes: Precise \& examples  (English)}
    \label{tab:gpt4_p4}
    \includegraphics[width=80mm]{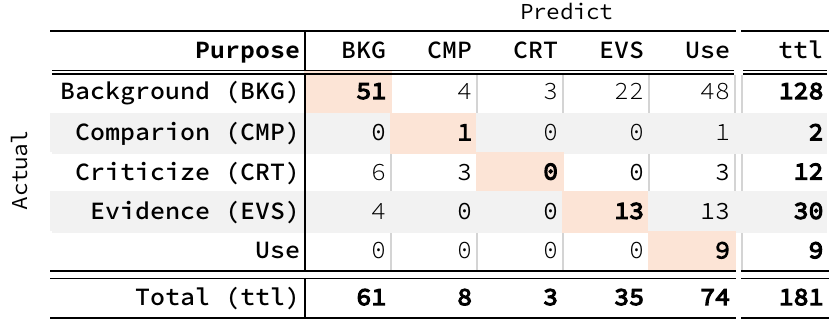}
\end{table}

In the article, we used the annotation results from the prompt (Precise, EN) with the highest consistency in order to evaluate the predictive performance of \texttt{gpt-3.5-turbo}.
In the follow-up study described above, \tabref{tab:gpt4_p3} shows the result for the same prompt.

While the prediction results using GPT3 were concentrated on "Background (BKG)" and "Evidence (EVS)", the number of other classes increased relatively more when GPT4 was used.
However, the performance is rather degraded due to the skewed distribution of the gold standard in the first place.
Compared to Table 9 
in the article, which summarizes the results from \texttt{gpt-3.5-turbo}, \tabref{tab:gpt4_p3} shows a decrease in the number of predicted BKG, which may be the reason for the lower performance.
On the other hand, the imbalance in the annotation results has been improved, resulting in some cases where "Criticize (CRT)" and "Use" were correctly predicted.
In particular, none were estimated to be Use when \texttt{gpt-3.5-turbo} was used, but their number increased significantly with \tabref{tab:gpt4_p3}.

From the above, it can be seen that the annotation results vary widely depending on the LLM model used.
However, in any case, the performance was not sufficient, suggesting that human annotation is still necessary for citation context analysis.

\subsection{Citation sentiment}

As well as citation purposes, the annotation results for citation sentiment are summarized in \tabref{tab:gpt4_s1} through \tabref{tab:gpt4_s3}.

\begin{table}[htbp]
    \centering
    \caption{Citation sentiment: Simple (English)}
    \label{tab:gpt4_s1}
    \includegraphics[width=80mm]{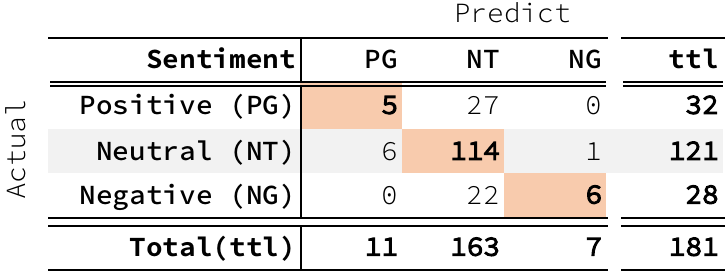}
\end{table}

\begin{table}[htbp]
    \centering
    \caption{Citation sentiment: Basic (English)}
    \label{tab:gpt4_s2}
    \includegraphics[width=80mm]{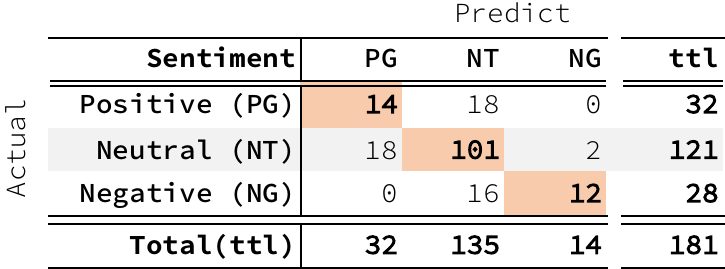}
\end{table}

\begin{table}[htbp]
    \centering
    \caption{Citation sentiment: Precise (English)}
    \label{tab:gpt4_s3}
    \includegraphics[width=80mm]{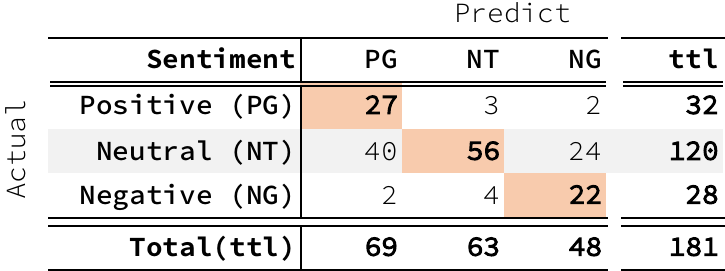}
\end{table}

In the article, we used the annotation results from the prompt (Simple, EN) with the highest consistency in order to evaluate the predictive performance of \texttt{gpt-3.5-turbo}.
In the follow-up study described above, 
\tabref{tab:gpt4_s1} shows the result for the same prompt.
The general trends shown in \tabref{tab:gpt4_s1} are similar to those in 
Table 10 
in the article, which suggests that there is no significant difference between \texttt{gpt-3.5-turbo} and \tabref{tab:gpt4_s1} other than a slight decrease in performance of the latter.

\end{document}